# Strain-Polarization Coupling in the Low-Dimensional Van der Waals Ferrielectrics

Anna N. Morozovska
*Institute of Physics*
*National Academy of Sciences of Ukraine*
Kyiv, Ukraine
anna.n.morozovska@gmail.com

Hanna V. Shevliakova
*Department of Microelectronics*
*National Technical University of Ukraine "Igor Sikorsky Kyiv Polytechnic Institute"*
Kyiv, Ukraine
shevliakova.hv@gmail.com

Liubomyr M. Korolevych
*Department of Microelectronics*
*National Technical University of Ukraine "Igor Sikorsky Kyiv Polytechnic Institute"*
Kyiv, Ukraine
korolevych.lyubomyr@gmail.com

Victoria Khist
*Department of General Physics*
*National Technical University of Ukraine "Igor Sikorsky Kyiv Polytechnic Institute"*
*Department of Physics of Meso- and Nanocrystalline Magnetic Structures*
*Institute of Magnetism*
*NAS of Ukraine and MES of Ukraine*
Kyiv, Ukraine
khist2012@gmail.com

Yulian M. Vysochanskii
*Institute of Solid State Physics and Chemistry,*
*Uzhhorod University, Uzhhorod, Ukraine*
vysochanskii@gmail.com

Eugene A. Eliseev
*Institute for Problems of Materials Science*
*National Academy of Sciences of Ukraine*
Kyiv, Ukraine
eugene.a.eliseev@gmail.com

*Abstract* — Using the Landau-Ginzburg-Devonshire phenomenological approach we explore the strain-polarization coupling in the low-dimensional van der Waals ferrielectrics. We evolve the analytical model of the piezoelectric susceptibility of the material in response to the periodic strain modulation, such as caused by a surface acoustic wave. Numerical calculations are performed for recently discovered and poorly studied ultrathin layers of vdW ferrielectric $CuInP_2S_6$, which are very promising candidates for advanced nanoelectronics. We obtained that the temperature dependences of the dielectric and piezoelectric susceptibilities, and elastic compliance of the ultrathin $CuInP_2S_6$ layer have a sharp maximum at the temperature of $CuInP_2S_6$ paraelectric-ferrielectric phase transition near 320 K. The magnitudes of the dielectric and piezoelectric susceptibilities, and elastic compliance depend significantly on the modulation period of the surface acoustic wave. Obtained results explore the potential of ultrathin $CuInP_2S_6$ layers for application in piezoelectric and straintronic devices.

## I. Introduction

Two-dimensional van der Waals (vdW) materials, the first representative of which is graphene, have long been of interest to researchers [1]. A unique feature of vdW materials is the strong interatomic bonds within the monolayer, and weaker van der Waals forces between the layers. Thanks to this, they can be divided into single-layer structures. Due to the van der Waals forces between the layers, these materials are convenient for the creation of heterostructures based on them since the formation due to van der Waals forces does not require agreement in lattice constants. VdW materials have significant differences between out-of-plane and in-plane properties, and the properties of single-layers, few-layers and their bulk materials can be principally different. For example, a wide tunability of polar and semiconducting properties of 1-, 2- or 3-layer structures of vdW low-dimensional transition metal dichalcogenides (such as $MoS_2$, $WS_2$, $ReS_2$, $MoSe_2$) [2,3] and their Janus-compounds (such as $MoSSe$, $BiTeI$) [4,5] made these unique materials most important for fundamental and applied physical research.

A very interesting representative of vdW materials is a Copper indium thiophosphate (CIPS), with chemical formula $CuInP_2S_6$, not only because of its vdW structure, but also the presence of reliable out-of-plane ferroelectricity at temperatures close to the room [6]. Its unique ferroelectric properties are largely induced by Cu off-centering (Fig. 1) [7]. Also, the negative capacitance of CIPS was detected, which allows to improve the control of the current in the channel of the field-effect transistors (FETs), thereby reducing power consumption and heating [8]. Therefore, due to the variety of its unique properties, CIPS has a broad range of applications in ferroelectric non-volatile random-access memories (FRAM) [9] and ferroelectric FETs [10]. The presence of spontaneous polarization immediately arouses interest in the study of other polar properties of these materials, such as piezoelectric [11,12], electrocaloric and pyroelectric [13] effects, thanks to which they can be used in pyroelectric nanogenerators [14] ultrasound transducers and solid-state cooling devices [15], as well as in piezoelectric and straintronic devices [11].

Despite the fact that many works are devoted to the study of CIPS, many aspects of its piezoelectric and elastic properties are not sufficiently studied and require additional analysis. The knowledge gap motivated this study, devoted to the Landau-Ginzburg-Devonshire (LGD) phenomenological description of the strain-polarization coupling in the low-dimensional vdW ferrielectrics. Below we present the analytical model of the piezoelectric susceptibility of the vdW material in response to the periodic strain modulation, such as caused by a surface acoustic wave (SAW). Numerical calculations are performed for a CIPS layer.

## II. Landau-Ginzburg-Devonshire Approach

The considered structure of CIPS (a-b) and the schematic model of metal-ferroelectric-semiconductor FET (MFS-FET) (c) is shown in Fig. 1.

The bulk density of the free energy $F$ that depends on polarization component $P$ and strain component $u$, and their gradients, has the following form:

$$F = \frac{\alpha}{2}P^2 + \frac{\beta}{4}P^4 + \frac{\gamma}{6}P^6 + \frac{\delta}{8}P^8 + \frac{g}{2}\left(\frac{\partial P}{\partial x}\right)^2 - quP^2 - zuP^4 - \frac{f}{2}\left(P\frac{\partial u}{\partial x} - u\frac{\partial P}{\partial x}\right) + \frac{c}{2}u^2 - PE - NU. \qquad (1)$$

According to Landau approach, the coefficient $\alpha$ should linearly depend on the temperature $T$, namely $\alpha(T) = \alpha_T(T - T_C)$, where $T_C$ is the Curie temperature of a bulk ferrielectric. Other coefficients in the free energy (1) are supposed to be temperature independent. The condition $\delta \geq 0$ is required for the stability of the free energy at arbitrary polarization values. The positive gradient coefficient $g$ determines the magnitude of the polarization gradient energy. Coefficient $f$ is the component of the static flexo-coupling tensor. The elastic stiffness $c$ should be positive for the free energy stability at arbitrary strain values. The linear and nonlinear electrostriction coefficients, $q$ and $z$, can be positive or negative. The elastic displacement component $U$ is related with the strain $u$ as $u = \partial U/\partial x$. $N$ is the bulk density of external mechanical force (e.g., periodic adhesion force induced by a patterned substrate).

$E$ is an external electric field in the free energy (1) and corresponding electrostatic contribution is $PE$. We did not include another electrostatic contribution, $PE_d/2$, in the free energy, because we assume that the depolarization field $E_d$ is small for the out-of-plane polarization orientation, e.g., due to the high conductivity of CIPS, or absent for the in-plane polarization orientation. One can neglect $E_d$ for the transverse fluctuations of polarization, which we consider hereinafter, and also assume that the longitudinal fluctuations of polarization are much smaller as being suppressed by the depolarization field.

The explicit form of the Euler-Lagrange type equations, $\delta F/\delta U = N$ and $\delta F/\delta P = 0$, are:

$$-c\frac{\partial^2 U}{\partial x^2} - f\frac{\partial^2 P}{\partial x^2} + 2qP\frac{\partial P}{\partial x} + 4zP^3\frac{\partial P}{\partial x} = N, \qquad (2a)$$

$$\alpha P + \beta P^3 + \gamma P^5 + \delta P^7 - g\frac{\partial^2 P}{\partial x^2} - f\frac{\partial^2 U}{\partial x^2} - 2qP\frac{\partial U}{\partial x} - 4zP^3\frac{\partial U}{\partial x} = E. \qquad (2b)$$

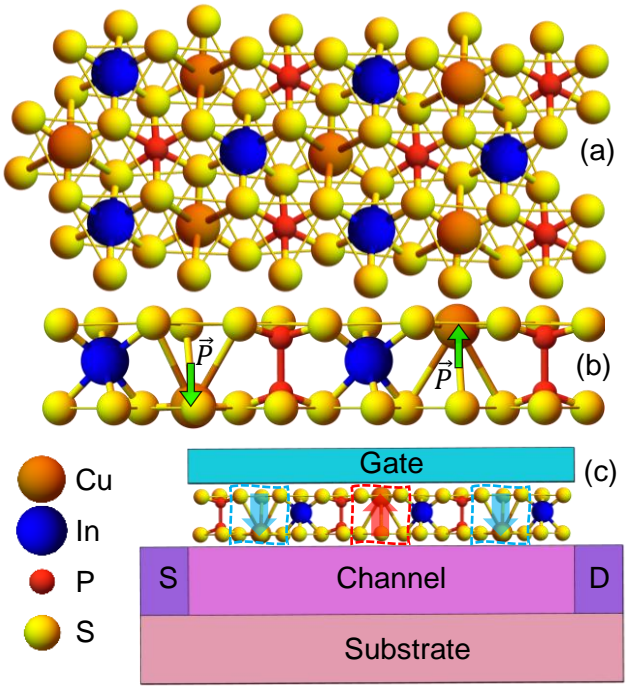

Fig. 1. Top (a) and side (b) view of CuInP$_2$S$_6$ (CIPS) structure and MFS-FET transistor with a CIPS ferroelectric gate (c). Thanks to CIPS, the channel can be made of wdW material in semiconductor phase, e.g. LD-TMDs.

The integral Fourier transform for polarization is:

$$P = P_S + \int dk \exp(ikx) \, \tilde{P},$$

where $P_S$ is the spontaneous polarization, and $k$ is the wave vector of the SAW.

The integral Fourier transform for elastic displacement is:

$$U = (u_S + u_m)x + \int dk \exp(ikx) \, \tilde{U},$$

where $u_S = \frac{1}{c}\left(qP_S^2 + zP_S^4\right)$ is the electrostriction-induced spontaneous strain, and $u_m$ is the misfit strain. In a general case the expression for $u_S$ follows from the equation of state, $\delta F/\delta u = \sigma$.

The integral Fourier transform for perturbating electric field is:

$$E = \int dk \exp(ikx) \, \tilde{E},$$

and the integral Fourier transform for perturbation elastic force density:

$$N = \int dk \exp(ikx)\, \tilde{N}.$$

Using the above Fourier transformations in the linearized Euler-Lagrange (2), we obtain the equations for the Fourier images of polarization $\tilde{P}$ and displacement $\tilde{U}$:

$$\left(\alpha - 2q(u_S + u_m) + 3(\beta - 4z(u_S + u_m))\right)P_S^2 + 5\gamma P_S^4 + +7\delta P_S^6 + gk^2)\tilde{P} + (fk^2 - 2iqkP_S - 4izkP_S^3)\tilde{U} = \tilde{E}. \tag{3b}$$

The linearized solution of the linearized (3) for the polarization has the form:

$$\tilde{P} = \tilde{\chi}(k)\varepsilon_0 \tilde{E} + \tilde{\eta}(k)\tilde{N}. \tag{4}$$

The linear susceptibilities, $\tilde{\chi}(k)$ and $\tilde{\eta}(k)$, introduced in (4), are given by the expressions:

$$\tilde{\chi}(k) = \frac{ck^2}{(\alpha_S + gk^2)ck^2 - (fk^2)^2 - (2qP_S + 4zP_S^3)^2 k^2},$$

$$\tilde{\eta}(k) = \frac{-(fk^2 - 2iqkP_S - 4izkP_S^3)}{(\alpha_S + gk^2)ck^2 - (fk^2)^2 - (2qP_S + 4zP_S^3)^2 k^2}.$$

Here the positive temperature-dependent dielectric stiffness is introduced:

$$\alpha_S(T) = \alpha(T) - 2q(u_S + u_m) + \\ + 3(\beta - 4z(u_S + u_m))P_S^2(T) + 5\gamma P_S^4(T) + 7\delta P_S^6.$$

Elastic displacement acquires the explicit form:

$$\tilde{U} = \tilde{\lambda}(k)\tilde{E} + \tilde{\vartheta}(k)\tilde{N}. \tag{5}$$

The linear compliances, $\tilde{\lambda}(k)$ and $\tilde{\vartheta}(k)$, introduced in (5), are given by the expressions:

$$\tilde{\lambda}(k) \equiv \tilde{\eta}^*(k) = \\ = \frac{-(fk^2 + 2iqkP_S + 4izkP_S^3)}{(\alpha_S + gk^2)ck^2 - (fk^2)^2 - (2qP_S + 4zP_S^3)^2 k^2},$$

$$\tilde{\vartheta}(k) = \frac{\alpha_S + gk^2}{(\alpha_S + gk^2)ck^2 - (fk^2)^2 - (2qP_S + 4zP_S^3)^2 k^2}.$$

The singularity of the generalized susceptibility corresponds to zero points of the denominator in (3). From the condition we derived the characteristic equation for the phase transition point:

$$\alpha_S(T)c + (cg - f^2)k^2 - [2qP_S(T) + zkP_S^3(T)]^2 = 0. \tag{6}$$

## III. RESULTS AND DISCUSSION

Fig. 2 and Fig. 3 illustrate the temperature dependence of key material properties in the $k$-domain for the CIPS. These properties are represented by the linearized solution for effective dielectric susceptibility $\tilde{\chi}(k)$, electromechanical

$$ck^2\tilde{U} + (fk^2 + 2ikqP_S + 4ikzP_S^3)\tilde{P} = \tilde{N}, \tag{3a}$$

coupling coefficient $\tilde{\eta}(k)$, and the effective elastic compliance $\tilde{\vartheta}(k)$. The modulation period (k) is varied to analyze its influence. The LGD CIPS parameters used in the calculations are listed in TABLE I.

Values of $\tilde{P}$ and $\tilde{U}$ are calculated for $\tilde{E} = 0$ from (4) and (5), respectively. The value of the critical temperature, $T_{tr}$, where the transition from the paraelectric to the ferrielectric phase occurs, and where the physical properties have peculiarities, is calculated from (6).

The observed temperature dependence for all three properties ($\tilde{\chi}(k)$, $\tilde{\eta}(k)$ and $\tilde{\vartheta}(k)$) is rather weak almost on the entire investigated temperature range. Significant changes are observed only around the critical temperature, $T_{tr} = 320$ K.

TABLE I.   LGD PARAMETERS FOR A BULK FERROELECTRIC CuInP$_2$S$_6$ AT FIXED STRAINS TAKEN FROM [11–13]

| Parameter | Value |
|---|---|
| $\varepsilon_b$ | 9 |
| $\alpha_T$ (C$^{-2}$·m J/K) | $1.64067 \times 10^7$ |
| $T_C$ (K) | 292.67 |
| $\beta$ (C$^{-4}$·m$^5$J) | $8.43 \cdot 10^{12}(1. - 0.00239\,T + 2.28 \cdot 10^{-6}T^2)$ |
| $\gamma$ (C$^{-6}$·m$^9$J) | $-1.67283 \cdot 10^{16}(1 - 0.00249\,T + 3.389 \times 10^{-6}\,T^2)$ |
| $\delta$ (C$^{-8}$·m$^{13}$J) | $9.824 \cdot 10^{18}(1 - 0.00127\,T + 4.0999 \cdot 10^{-6}T^2)$ |
| $q_{i3}$ (J C$^{-2}$·m) | $q_{13} = 1.4879 \cdot 10^{11}(1 - 0.00206\,T)$, $q_{23} = 1.0603 \cdot 10^{11}(1 - 0.00203\,T)$, $q_{33} = -4.0334 \cdot 10^{11}(1 - 0.00188\,T)$ |
| $z_{i33}$ (C$^{-4}$·m$^5$J) | $z_{133} = -1.414 \cdot 10^{14}(1 - 0.00099\,T)$, $z_{233} = -0.774 \cdot 10^{14}(1 - 0.00146\,T)$, $z_{333} = 1.181 \cdot 10^{14}(1 - 0.00069\,T)$ |
| $s_{ij}$ (Pa$^{-1}$) $c_{ij}$ (Pa) | $s_{11} = 1.510 \cdot 10^{-11}$, $s_{12} = 0.183 \cdot 10^{-11}$, $c_{11} = 6.803 \cdot 10^{10}$, $c_{12} = -7.364 \cdot 10^9$ |
| $g_{33ij}$ (J m$^3$/C$^2$) | $g \cong (0.5 - 2.0) \times 10^{-9}$ |

The graphs in Fig. 2 show that the effective dielectric susceptibility $\tilde{\chi}(k)$, real parts of the electromechanical coupling coefficient $\tilde{\eta}(k)$ monotonically increase in the interval from 0 to 320 K. At the transition point, $T_{tr}$, they reach the maximal value and begins to decrease slightly and monotonically. For all calculated values of the stress modulation period, $2\pi/k$, from 2.5 nm to 100 nm, the function form of the graphs is the same.

The temperature dependences of the imaginary part of the electromechanical coupling coefficient $\tilde{\eta}(k)$ represent a smooth rising lines, the function reaches its maximum value at temperature $T = T_{tr}$, which is slightly $k$-dependent.

The temperature dependences of the effective elastic compliance $\tilde{\vartheta}(k)$, shown in Fig. 2d, are parallel almost horizontal lines, which have a small jump at the transition temperature $T_{tr}$ for all values of the k-modulation period from 2.5 nm to 100 nm.

The dependences of effective dielectric susceptibility $\tilde{\chi}(k)$, real and imaginary parts of electromechanical coupling coefficient $\tilde{\eta}(k)$, and effective elastic compliance $\tilde{\vartheta}(k)$ on the temperature $T$ and modulation period $2\pi/k$ are shown in Fig. 3. Here the peculiarities at $T = T_{tr}$ are clearly visible.

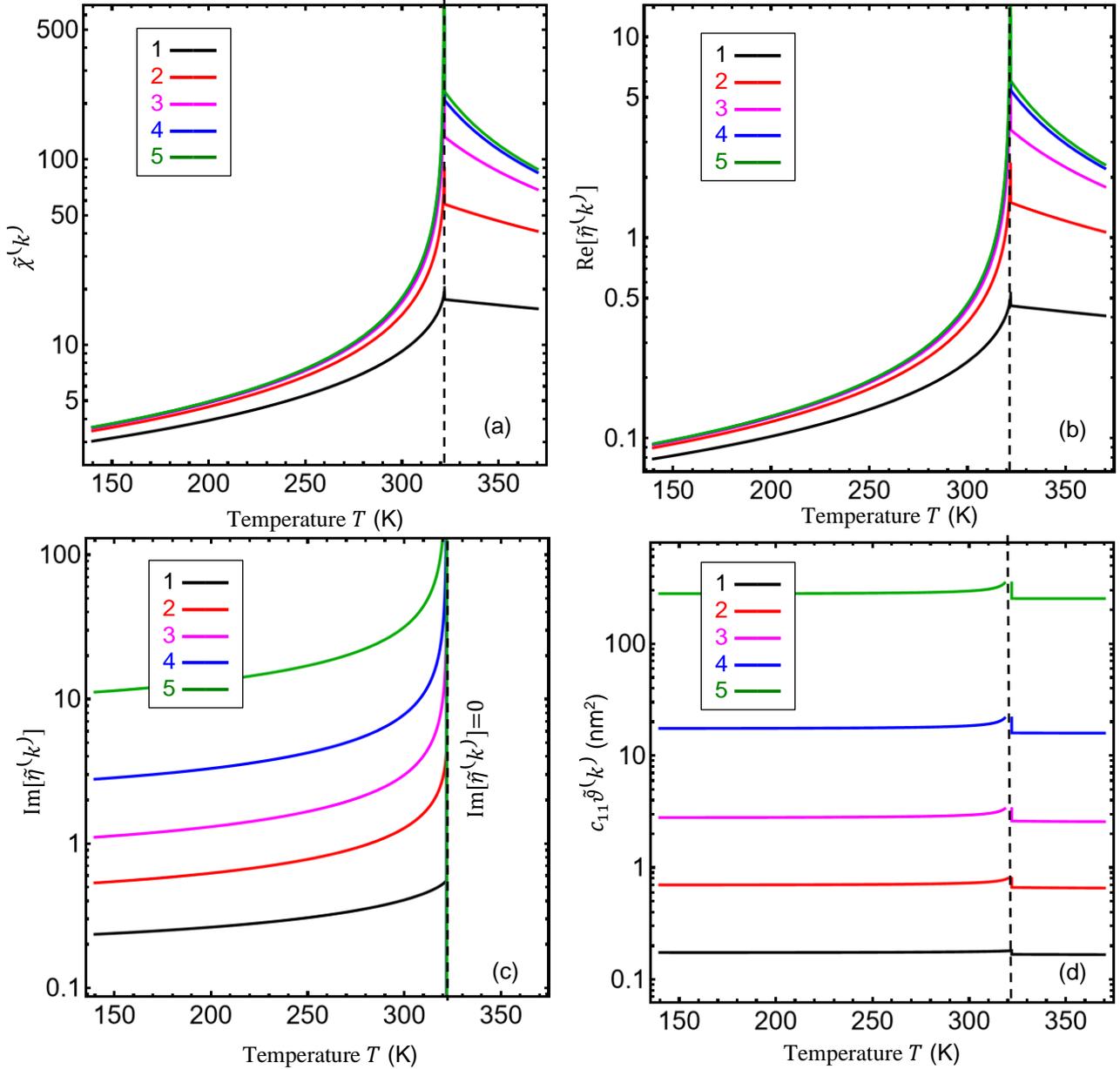

Fig. 2. Temperature dependences of the effective dielectric susceptibility $\tilde{\chi}(k)$ (a), real (b) and imaginary (c) parts of the electromechanical coupling coefficient $\tilde{\eta}(k)$, and the effective elastic compliance $\tilde{\vartheta}(k)$ (d) calculated for several values of the stress modulation period $2\pi/k = 2.5, 5, 10, 25$ and 100 nm (curves 1, 2, 3,4 and 5, respectively). Part a) is adapted from the Supplementary Materials to [11] for completeness.

## IV. CONCLUSIONS

Using the LGD phenomenological approach we studied the strain-polarization coupling in the low-dimensional van der Waals ferrielectrics. We evolve the analytical expressions of the piezoelectric susceptibility of the material in response to the periodic strain modulation, such as caused by a SAW. Numerical calculations are performed for the vdW ferrielectric CIPS, which is a very promising candidate for advanced nanoelectronics. It was found that the dielectric susceptibility, elastic compliance and the real part of piezoelectric susceptibility reach their maxima at the transition temperature $T_{tr} \approx 320$ K. Also, the simulation results indicate a significant dependence of the strain-polarization coupling parameters on the SAW modulation period. This result opens the way for CIPS layers application in piezoelectric and straintronic devices.

### DATA AVAILABILITY STATEMENT

Numerical results presented in the work are obtained and visualized using a specialized software, Mathematica 14.0. The Mathematica notebooks, which contain the codes, are available upon reasonable request.

### ACKNOWLEDGMENT

A.N.M. work is supported by the Ministry of Science and Education of Ukraine (grant № PH/ 23 - 2023, "Influence of size effects on the electrophysical properties of graphene-ferroelectric nanostructures") and support from the Horizon Europe Framework Programme (HORIZON-TMA-MSCA-SE), project № 101131229, Piezoelectricity in 2D-materials: materials, modelling, and applications (PIEZO 2D).

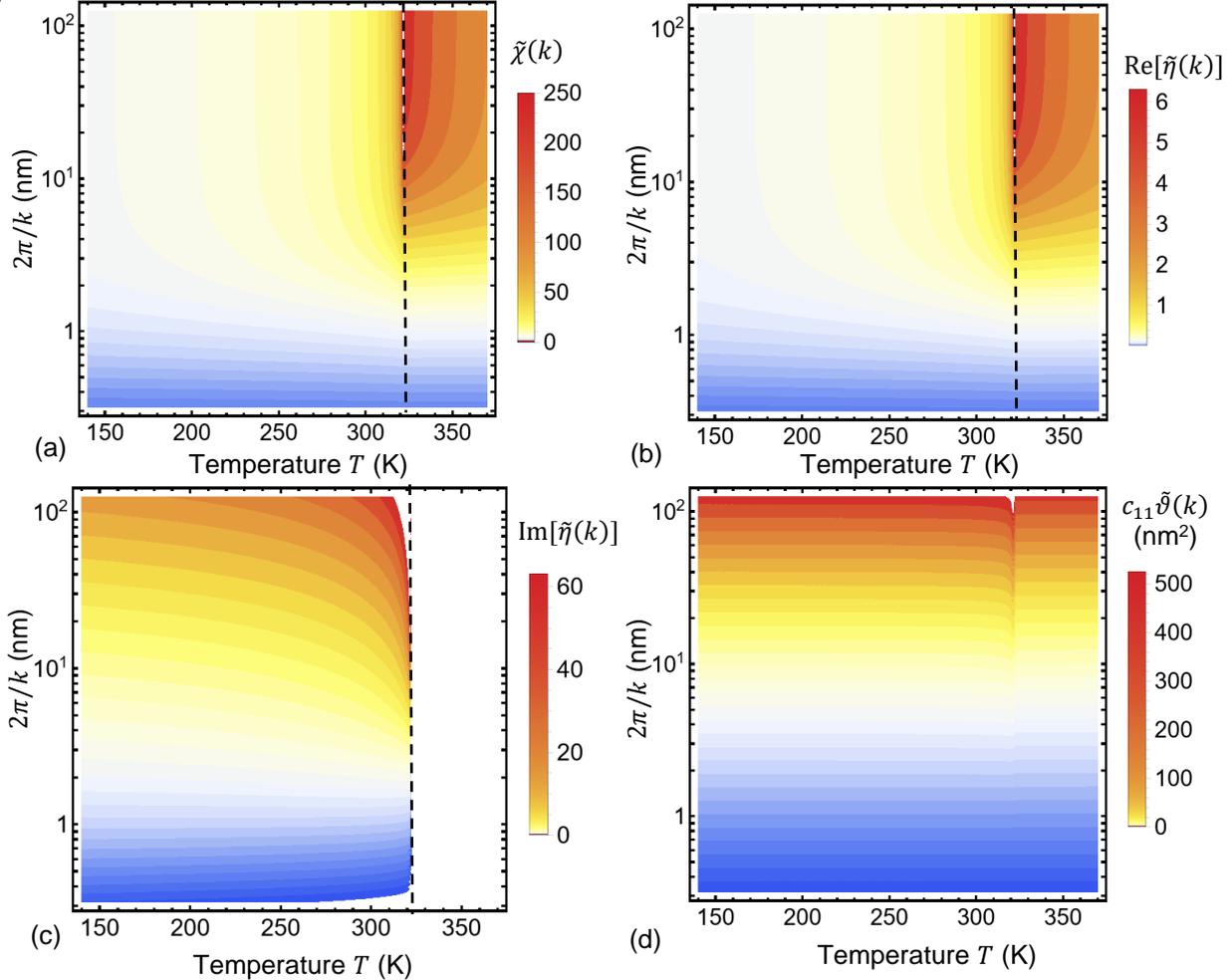

Fig. 3. The dependences of effective dielectric susceptibility $\tilde{\chi}(k)$ (a), real (b) and imaginary (c) parts of electromechanical coupling coefficient $\tilde{\eta}(k)$, and effective elastic compliance $\tilde{\vartheta}(k)$ (d) on the temperature $T$ and modulation period $2\pi/k$. Part (a) is adapted from the Supplementary Materials to [11] for completeness.

## REFERENCES

[1] K. S. Novoselov, A. Mishchenko, A. Carvalho, and A. H. Castro Neto, "2D materials and van der Waals heterostructures," *Science (1979)*, vol. 353, no. 6298, Jul. 2016, doi: 10.1126/science.aac9439.

[2] S. Kang, S. Kim, S. Jeon, W. Jang, D. Seol *et al.*, "Atomic-scale symmetry breaking for out-of-plane piezoelectricity in two-dimensional transition metal dichalcogenides," *Nano Energy*, vol. 58, pp. 57–62, Apr. 2019, doi: 10.1016/j.nanoen.2019.01.025.

[3] R. Grasset, R. Grasset, Y. Gallais, A. Sacuto, M. Cazayous, S. Mañas-Valero *et al.*, "Pressure-induced collapse of the charge density wave and Higgs mode visibility in 2H−TaS2," *Phys Rev Lett*, vol. 122, no. 12, p. 127001, Mar. 2019, doi: 10.1103/PhysRevLett.122.127001.

[4] Y. Qi, W. Shi, P. G. Naumov, N. Kumar, R. Sankar *et al.*, "Topological quantum phase transition and superconductivity induced by pressure in the bismuth tellurohalide BiTeI," *Advanced Materials*, vol. 29, no. 18, p. 1605965, May 2017, doi: 10.1002/adma.201605965.

[5] L. Dong, J. Lou, and V. B. Shenoy, "Large in-plane and vertical piezoelectricity in Janus transition metal dichalcogenides," *ACS Nano*, vol. 11, no. 8, pp. 8242–8248, Aug. 2017, doi: 10.1021/acsnano.7b03313.

[6] S. Zhou, L. You, H. Zhou, Y. Pu, Z. Gui, and J. Wang, "Van der Waals layered ferroelectric CuInP2S6: Physical properties and device applications," *Front Phys (Beijing)*, vol. 16, no. 1, p. 13301, Feb. 2021, doi: 10.1007/s11467-020-0986-0.

[7] F. Liu, L. You, K. L. Seyler, X. Li, P. Yu *et al.*, "Room-temperature ferroelectricity in CuInP2S6 ultrathin flakes," *Nat Commun*, vol. 7, no. 1, p. 12357, Aug. 2016, doi: 10.1038/ncomms12357.

[8] S. M. Neumayer, L. Tao, A. O'Hara, M. A. Susner, M. A. McGuire *et al.*, "The concept of negative Capacitance in ionically conductive van der Waals ferroelectrics," *Adv Energy Mater*, vol. 10, no. 39, Oct. 2020, doi: 10.1002/aenm.202001726.

[9] X. Jiang, X. Hu, J. Bian, K. Zhang, L. Chen *et al.*, "Ferroelectric field-effect transistors based on WSe2/CuInP2S6 heterostructures for memory Applications," *ACS Appl Electron Mater*, vol. 3, no. 11, pp. 4711–4717, Nov. 2021, doi: 10.1021/acsaelm.1c00492.

[10] M. Si, P.-Y. Liao, G. Qiu, Y. Duan, and P. D. Ye, "Ferroelectric field-effect transistors based on MoS2 and CuInP2S6 two-dimensional van der Waals heterostructure," *ACS Nano*, vol. 12, no. 7, pp. 6700–6705, Jul. 2018, doi: 10.1021/acsnano.8b01810.

[11] Y. Liu, A. N. Morozovska, A. Ghosh, K. P. Kelley, E. A. Eliseev *et al.*, "Stress and curvature effects in layered 2D ferroelectric CuInP2S6," *ACS Nano*, vol. 17, no. 21, pp. 22004–22014, Nov. 2023, doi: 10.1021/acsnano.3c08603.

[12] A. N. Morozovska, E. A. Eliseev, Y. Liu, K. P. Kelley, A. Ghosh *et al.*, "Bending-induced isostructural transitions in ultrathin layers of van der Waals ferrielectrics," *Acta Mater*, vol. 263, p. 119519, Jan. 2024, doi: 10.1016/j.actamat.2023.119519.

[13] A. N. Morozovska, E. A. Eliseev, L. P. Yurchenko, V. V. Laguta, Y. Liu *et al.*, "The strain-induced transitions of the piezoelectric, pyroelectric, and electrocaloric properties of the CuInP2S6 films," *AIP Adv*, vol. 13, no. 12, Dec. 2023, doi: 10.1063/5.0178854.

[14] W. F. Io, M.-C. Wong, S.-Y. Pang, Y. Zhao, R. Ding *et al.*, "Strong piezoelectric response in layered CuInP2S6 nanosheets for piezoelectric nanogenerators," *Nano Energy*, vol. 99, p. 107371, Aug. 2022, doi: 10.1016/j.nanoen.2022.107371.

[15] M. Si, A. K. Saha, P.-Y. Liao, S. Gao, S. M. Neumayer, J. Jian *et al.*, "Room-temperature electrocaloric effect in layered ferroelectric CuInP$_2$S$_6$ for solid-state refrigeration," *ACS Nano*, vol. 13, no. 8, pp. 8760–8765, Aug. 2019, doi: 10.1021/acsnano.9b01491.